\documentclass[]{ieeeaccess}

\usepackage{cite}
\usepackage{amsmath,amssymb,amsfonts}
\usepackage{algorithmic}
\usepackage{graphicx}
\usepackage{textcomp}
\usepackage{mathptmx}
\usepackage{bm}
\usepackage[english]{babel}
\usepackage{url}
\usepackage{color}
\usepackage{caption2}
\usepackage{euscript}
\usepackage{mathptmx}
\usepackage{algorithm}
\usepackage{verbatim}
\usepackage{multirow}
\usepackage{setspace}
\usepackage[tight,footnotesize]{subfigure}
\usepackage{psfrag}
\usepackage{latexsym}
\usepackage{graphicx}
\usepackage{textcomp}
\usepackage{multirow}

\newtheorem{lemma}{Lemma}

\newcommand{\Es}{{\mathbb{E}}}          
\newcommand{\rank}{{\text{rank}}}
\newcommand{\diag}{{\text{diag}}}
\newcommand{\trace}{{\text{tr}}}

\newcommand{\I}{\bm{I}}

\newcommand{\Nb}{{N_{\text{B}}}}

\newcommand{\Ns}{{N_{\text{S}}}}

\newcommand{\Nc}{{N_{\text{C}}}}
\newcommand{\Nr}{{N_{\text{R}}}}
\newcommand{\Nd}{{N_{\text{D}}}}

\newcommand{\z}{\bm{z}}

\newcommand{\A}{\bm{A}}

\newcommand{\R}{\bm{R}}

\newcommand{\E}{\bm{E}}

\newcommand{\C}{\bm{C}}
\newcommand{\V}{\bm{V}}
\newcommand{\D}{\bm{D}}
\newcommand{\Dmmse}{\bm{D}_{\text{mmse}}}
\newcommand{\Q}{\bm{Q}}
\newcommand{\U}{\bm{U}}

\newcommand{\w}{\bm{w}}

\newcommand{\n}{\bm{n}}
\renewcommand{\b}{\bm{b}}
\renewcommand{\r}{\bm{r}}
\renewcommand{\a}{\bm{a}}

\newcommand{\x}{\bm{x}}

\newcommand{\F}{\bm{F}}

\renewcommand{\H}{\bm{H}}
\newcommand{\G}{\bm{G}}
\renewcommand{\v}{\bm{v}}

\newcommand{\Cset}{\mathbb{C}}
\newcommand{\Rset}{\mathbb{R}}

\newcommand{\Ptot}{\mathcal{P}_{\text{tot}}}

\newcommand{\eqdef}{\triangleq}

\renewcommand{\det}{{\mathrm{det}}}
\newcommand{\herm}{\text{H}}
\newcommand{\trasp}{\text{T}}

\newcommand{\istar}{i^\star}

\newcommand{\Psource}{{\mathcal{P}}_{\text{S}}}

\newcommand{\GlobalPrelay}{{\mathcal{P}}_{\text{D}}}
\newcommand{\GlobalPrelayup}{\widetilde{\mathcal{P}}_{\text{D}}}

\newcommand{\Rvv}{\bm{K}_{\v\v}}
\newcommand{\Rzz}{\bm{K}_{\z\z}}

\def\bdm#1\edm{\begin{displaymath}#1\end{displaymath}}
\def\be#1\ee{\begin{equation}#1\end{equation}}
\def\barr#1\earr{\begin{align}#1\end{align}}

\newcommand{\IeeeTIT}{\textit{IEEE Trans.\ Inf.\ Theory\/}}
\newcommand{\IeeeTSP}{\textit{IEEE Trans.\ Signal Process.\/}}
\newcommand{\IeeeTCOMM}{\textit{IEEE Trans.\ Commun.\/}}
\newcommand{\IeeeCOMMLETT}{\textit{IEEE Commun.\ Lett.\/}}

\newcommand{\IeeeTWC}{\textit{IEEE Trans.\ Wireless Commun.\/}}
\newcommand{\IeeeJSAC}{\textit{IEEE J.\ Select.\ Areas Commun.\/}}
\newcommand{\IeeeTVT}{\textit{IEEE Trans.\ Veh.\ Technol.\/}}

\newcommand{\IeeeSJ}{\textit{IEEE Syst.\ J.\/}}

\newcommand{\EurasipJASP}{\textit{EURASIP J.\ Advances Signal Process.\/}}

\newcommand{\IeeeTIFS}{\textit{IEEE Trans.\ Inf.\ Forensics Security}}
\newcommand{\IeeeTCE}{\textit{IEEE Trans.\ Consumer Electron.\/}}

\begin{document}

\history{Date of publication xxxx yy, 2020, date of current version June 8, 2020.}
\doi{zzzzzzzzzzzzzzzzz}

\title{Design of cooperative MIMO wireless sensor networks with
partial channel state information
}

\author{
\uppercase{Donatella Darsena}\authorrefmark{1}, \IEEEmembership{Senior Member, IEEE},
\uppercase{Giacinto Gelli}\authorrefmark{2}, \IEEEmembership{Senior Member, IEEE} AND 
\uppercase{Francesco Verde}\authorrefmark{2}, \IEEEmembership{Senior Member, IEEE}
}
\address[1]{Department of Engineering, Parthenope University, Naples I-80143, Italy
(e-mail: darsena@uniparthenope.it)}
\address[2]{Department of Electrical Engineering and
Information Technology, University Federico II, Naples I-80125,
Italy [e-mail: (gelli, f.verde)@unina.it]}

\markboth
{Darsena \headeretal: Design of cooperative MIMO wireless sensor networks with
partial channel state information}{}

\corresp{Corresponding author: Francesco Verde (e-mail: f.verde@unina.it).}

\begin{abstract}

Wireless sensor networks (WSNs) play a key role 
in automation and consumer electronics applications.
This paper deals with joint design of the source precoder,
relaying matrices, and destination equalizer in a multiple-relay
amplify-and-forward (AF) cooperative multiple-input
multiple-output (MIMO) WSN, when partial
channel-state information (CSI) is available in the network.
In particular, the considered approach assumes knowledge of
instantaneous CSI of the first-hop channels and 
statistical CSI of the second-hop channels.
In such a scenario, compared to the case when instantaneous CSI of
both the first- and second-hop channels 
is exploited, existing network designs 
exhibit a significant performance degradation.
Relying on a relaxed minimum-mean-square-error (MMSE)
criterion, we show that strategies based on potential activation of
all antennas belonging to all relays 
lead to mathematically intractable 
optimization problems.
Therefore, we develop a new joint relay-and-antenna 
selection procedure, which determines the best subset 
of the available antennas possibly 
belonging to different
relays. 
Monte Carlo simulations show that, compared to conventional relay selection strategies,
the proposed design offers a significant performance gain, 
outperforming also other
recently proposed 
relay/antenna selection schemes.

\end{abstract}


\titlepgskip=-15pt

\maketitle

\section{Introduction and system model}
\label{sec:intro}

\IEEEPARstart{W}{ith} 
the advent of massive Internet of Things and 
massive machine-type communications, especially in the domain of 
5G consumer electronics, there is a need to further enhance
physical-layer performance of \textit{ wireless sensor networks (WSNs)}.
In this respect, \textit{ amplify-and-forward (AF)} relaying
is an effective way
to improve transmission reliability over fading channels, by
taking advantage of the broadcast nature of wireless communications
\cite{Laneman2004,Yuan.2010}, especially when the network nodes are
equipped  with multiple-input multiple-output (MIMO) transceivers 
\cite{Goldsmith,Fand.2010,Nguyen}.

\IEEEpubidadjcol

We consider a one-way cooperative MIMO WSN aimed at
transmitting a symbol block $\b \in \Cset^{\Nb}$ 
from a source to a destination, with the assistance of 
$\Nc$ half-duplex relays.\footnote{
The fields of complex and numbers are
denoted with $\mathbb{C}$ and $\mathbb{R}$,
respectively;
matrices [vectors] are denoted
with upper [lower] case boldface
letters (e.g., $\A$ or $\a$);
the field of $m \times n$
complex [real] matrices is denoted
as $\Cset^{m \times n}$ [$\Rset^{m \times n}$],
with $\Cset^m$ [$\Rset^m$] used as a
shorthand for
$\Cset^{m \times 1}$ [$\Rset^{m \times 1}$];
the superscripts
$*$, $T$, $H$, $-1$, and $\dag$
denote the conjugate,
the transpose, the conjugate transpose,
the inverse, and the Moore-Penrose generalized
inverse of a matrix, respectively;
$\{\A \}_{ij}$
indicates the $(i+1,j+1)$th element of
$\A \in \Cset^{m \times n}$, with
$i \in \{0,1, \ldots, m-1\}$
and $j\in\{0,1,\ldots,n-1\}$;
$\bm{0}_{m} \in \Rset^m$, 
$\mathbf{O} \in \Rset^{m \times n}$,
and $\I_{m} \in \Rset^{m \times m}$
denote the null vector, the null matrix, and
the identity matrix, respectively;
$\trace(\A)$ denotes the trace  of $\A \in \Cset^{n \times n}$;
$\rank(\A)$ is the rank of $\A \in \Cset^{m \times n}$;
finally, the operator $\Es[\,\cdot\,]$ denotes ensemble averaging.
}
We assume that there is no direct link between the source and the destination,
due to high path loss values or obstructions, and we denote with
$\Ns$, $\Nr$, and $\Nd$, respectively,
the numbers of antennas at the source, relays, 
and destination.
The received signal at the destination can be expressed as
\be
\r =  \C \, \b + \v
\label{eq:ric}
\ee
where $\C \eqdef \G \, \F \, \H \, \F_0 \in \Cset^{\Nd \times \Nb}$
is the \textit{ dual-hop} channel matrix and $\v \eqdef \G \, \F \, \w + \n$ is the
equivalent noise vector at the destination.
The composite matrices 
\begin{eqnarray}
\H  & \eqdef [\H_1^{ \trasp}, \H_2^{\trasp}, \ldots,
\H_{\Nc}^{ \trasp}]^{\trasp} \in \Cset^{(\Nc \Nr) \times \Ns}
\\
\G  & \eqdef [\G_{1},
\G_{2}, \ldots, \G_{N_c}] \in \Cset^{\Nd \times (\Nc \Nr)}
\end{eqnarray}
collect the \textit{ first-} (backward) and \textit{ second-hop} (forward) 
MIMO channel coefficients of
all the relays, respectively, whereas
the diagonal blocks $\F_i \in \Cset^{\Nr \times \Nr}$ of
\be
\F  \eqdef \diag(\F_1, \F_2, \ldots, \F_{\Nc})
\ee
denote the \emph{relaying matrices}, and
$\F_0 \in \Cset^{\Ns \times \Nb}$ represents the \emph{source precoding matrix}.
Finally, $\w \in \Cset^{\Nc \Nr}$ and $\n \in \Cset^{\Nd}$ 
gather the noise samples at all the relays and  at the destination, respectively.
The vector $\r$
is subject to linear equalization at the destination through
the \emph{equalizing matrix} $\D \in \Cset^{\Nb \times \Nd}$,
hence yielding an estimate
$\hat{\b} \eqdef \D \, \r$ of the source block $\b$,
whose entries are then subject to minimum-distance (in the Euclidean sense) detection.
Increase in spectral efficiency can be obtained by considering 
\textit{ two-way relaying} \cite{Rankov}, which is based
on establishing bidirectional connections 
between two or more terminals
using one or several half-duplex relays.

To achieve the expected gains, 
channel state information (CSI) is required at the network nodes, 
i.e, source, AF relays, and destination.
\textit{ Full CSI (F-CSI)} is invoked in many papers
dealing with optimization of one-way
(see, e.g., \cite{Merched.2008,Toding.2012,Truong2013,Champagne2013, Lee2014,Ikhlef2014,Champagne2015,Darsena.2017,Darsena-jcn})
and two-way (see, e.g., \cite{Zhang,Yang})
cooperative MIMO networks.
Specifically, with reference to the system model \eqref{eq:ric},
F-CSI is tantamount to requiring:
(i) instantaneous knowledge of the first-hop channel matrix $\H$;
(ii) instantaneous knowledge of the second-hop channel matrix $\G$;
(iii) instantaneous knowledge of the dual-hop channel matrix $\C$.
While the dual-hop channel matrix $\C$ can be directly
estimated at the destination by training, 
separate acquisition of the first- and 
second-hop matrices $\H$ and $\G$
is more complicated to achieve, 
both in terms of communication resources and signal
overhead, especially in multiple-relay WSNs.
Moreover, since channel estimation errors occur in practical situations,
robust optimization designs are needed \cite{Chalise,Li},
which further complicate system deployment.
In resource-constrained WSNs, 
the use of \textit{ partial} CSI (P-CSI)
can extend network lifetime and reduce the complexity burden.

\emph{Relay selection} is a common strategy 
to reduce signaling overhead and system design 
complexity in single-input single-output (SISO) cooperative WSNs
\cite{Bletsas.2006,Zhao.2006,Jing.2009,Sun.2009}.
Design of SISO relay selection procedures
providing diversity gains -- even when
F-CSI is not available -- has been addressed
in \cite{Krikidis.2008,Yi.2007,Kim.2009,Costa.2009,Yi2007}.
Such methods rely on P-CSI, since
selection of the best relay is based only on
instantaneous knowledge of the source-to-relay channels.
However, the diversity order of the 
methods developed in these papers does not 
scale in the number of relays $\Nc$.
For SISO nodes, a P-CSI relay selection scheme
has been proposed in \cite{Chalise2013}, yielding full diversity
order $\Nc$.
However, besides the instantaneous 
knowledge of the source-to-relay channels,
such a method requires that the selected 
relay sends instantaneous CSI of the corresponding
source-to-relay channel to the
destination for optimal decoding. 
Moreover, the optimization problem in \cite{Chalise2013} 
does not admit a closed-form solution and is solved
by using a line search algorithm.

It has been shown in \cite{Chalise2012} that P-CSI relay selection 
approaches for MIMO nodes, 
based only on the instantaneous knowledge 
of $\H$, do not fully exploit the diversity arising 
from the presence of multiple relays. 
Besides instantaneous knowledge of $\H$, statistical CSI of the
second-hop matrix $\G$ is used in \cite{Darsena.2017, Danaee2012}
to perform relay/antenna selection for a MIMO AF cooperative network.
However, the solutions developed in
\cite{Darsena.2017, Danaee2012} still exhibit a significant performance degradation
compared to designs based on F-CSI.

In this paper, we present new optimization methods for
multiple-relay cooperative MIMO WSNs with P-CSI,
i.e., knowledge of the instantaneous value of $\H$ and the 
statistical properties of $\G$.
Our design does not rely on F-CSI as in
\cite{Merched.2008,Toding.2012,Truong2013,Champagne2013, Lee2014,Ikhlef2014,Champagne2015,Darsena-jcn,Zhang,Yang},
and needs the same amount of P-CSI exploited in 
\cite{Krikidis.2008,Yi.2007,Kim.2009,Costa.2009,Yi2007,Danaee2012}.
In this scenario, we consider a relaxed joint minimum-mean-square-error (MMSE)
optimization of the source precoder $\F_0$, the AF relaying matrices in $\F$,
and the destination equalizer $\D$, with a power constraint at the source
\cite{Palomar.2003} and \textit{ a sum-power constraint} at the relays \cite{Champagne2013}.
Specifically, capitalizing on our preliminary results \cite{Darsena.2017},
the novel contributions can be summarized as follows:

\begin{enumerate}

\item
We prove that the MMSE-based 
design attempting to activate all possible 
antennas of all relays leads to a
mathematically intractable optimization problem.

\item
We provide the proofs
of the results reported in \cite{Darsena.2017},
by enlightening that single relay selection 
\cite{Darsena.2017,Krikidis.2008,Yi.2007,Kim.2009,Costa.2009,Yi2007} 
is suboptimal in the considered P-CSI scenario.

\item
We develop a new joint antenna-and-relay selection 
algorithm, which is shown to significantly
outperform the relay/antenna selection approaches
\cite{Darsena.2017, Yi.2007,Danaee2012} in terms of
average symbol error probability (ASEP).

\end{enumerate}

The paper is organized as follows.
Section~\ref{sec:signal} introduces the basic assumptions
and discusses their practical implications.
The proposed designs are developed in Section~\ref{sec:theor-a}.
Section~\ref{sec:simulation} reports simulation
results in terms of ASEP, whereas Section~\ref{sec:concl} draws some conclusions.

\section{Basic assumptions and preliminaries}
\label{sec:signal}

The symbol block $\b$ in \eqref{eq:ric} is modeled as a 
circularly symmetric
complex random vector, with
$\Es[\b \, \b^{\herm}] = \I_{\Nb}$.
The entries of $\H$ and $\G$ are assumed to be
unit-variance circularly symmetric complex
Gaussian (CSCG) random variables.
The noise vectors $\w$ and $\n$
are modeled as mutually independent CSCG
random vectors, statistically independent of $(\b,\H,\G)$, with
$\Es[\w \, \w^{\herm}] = \I_{\Nc \Nr}$
and $\Es[\n \, \n^{\herm}] = \I_{\Nd}$, respectively.

Hereinafter, we assume that $\C$ in \eqref{eq:ric} and
the following conditional covariance matrix of $\v$, given $\G$,
\be
\Rvv \eqdef \Es[\v \, \v^\herm \, | \, \G] =\G \, \F \, \F^\herm \G^\herm +  \I_{\Nd}
\label{eq:Rvv}
\ee
have been previously
acquired at the destination during
a training session.
Under such assumptions, it is known (see, e.g., \cite{Palomar.2003}) that, for fixed
matrices $\F_0$ and $\F$,  the matrix $\D$ minimizing
the trace of the conditional mean square error (MSE) matrix
$\E(\F_0,\F,\D)   \eqdef \Es[(\hat{\b}-\b) \, (\hat{\b}-\b)^\herm \, | \, \H, \G]$, given
$\H$ and $\G$,  is the Wiener filter
\be
\Dmmse =  \C^\herm  (\C \, \C^\herm+\Rvv)^{-1} \: .
\ee

Optimization of $\F_0$ and $\F$ 
is carried out under the assumption that only
P-CSI is available at the source and the relays.
Specifically, the source and the relays
perfectly know the first-hop 
channel matrix $\H$, but the
$i$th relay has only knowledge 
of the second-order statistics (SOS)
of its own second-hop channel 
matrix $\G_i$.
These assumptions are justified since, 
in some systems, the relays may be 
able to exchange information
among themselves before transmission \cite{Ela.2014}.
In this case, knowledge of $\H$ at the relays 
is realistic \cite{Merched.2008,Toding.2012,Truong2013,Champagne2013, Lee2014,Ikhlef2014,Champagne2015,Darsena.2017, Bletsas.2006,Zhao.2006,Jing.2009,
Krikidis.2008,Kim.2009,Costa.2009,Yi2007,Chalise2012,Chalise2013}.
Moreover,  since the SOS of $\G_i$ vary much more slowly
than the instantaneous values of $\G_i$, 
the feedback overhead from the destination to the relays
is significantly reduced, compared to \cite{Chalise2013}.

\section{The proposed P-CSI-based design}
\label{sec:theor-a}

To obtain $\F_0$ and $\F$,
we minimize the statistical average 
(with respect to $\G$) of
the trace of the following matrix:
\be
\E(\F_0,\F) \eqdef  \E(\F_0,\F,\Dmmse) =
(\I_{\Nb}+\C^\herm \, \Rvv^{-1} \, \C)^{-1}
\ee
under suitable power constraints.
To this aim, we assume that:
{\bf a1)}  $\F_0$ is full-column rank, i.e., $\rank(\F_0)=\Nb \le \Ns$;
{\bf a2)} $\G  \F  \H$ is full-column rank, i.e., $\rank(\G  \F  \H) = \Ns \leq \Nd$.
It is noteworthy that assumption {\bf a2)} necessarily requires that
the matrices  $\F  \H$ and $\H$ are full-column rank, i.e.,
$\rank(\F \H)=\rank(\H)=\Ns \le \Nc \, \Nr$.
Such assumptions ensure that $\C$ is full-column rank as well.
Specifically, we consider the following optimization problem:
\begin{multline}
\min_{\F_0 ,\bm F}
\Es_{\G} \left\{\trace\left[ \left(\I_{\Nb}+\C^\herm  \, \Rvv^{-1}  \, \C\right)^{-1}\right]
\, \big | \,  \H \right\}
\,  \text{subject to (s.to)} \\
\trace(\F_0 \, \F_0^{\herm}) \leq \Psource \quad \text{and} \quad
\Es_{\G} \left [\trace(\G  \,\F  \,\Rzz  \,\F^\herm  \G^\herm)  \, |\,  \H \right]
\leq \GlobalPrelay
\label{eq:mat-def-1}
\end{multline}
where $\Rzz \eqdef \Es[\z  \,\z^\herm  \,|\,  \H]  = \H  \,\F_0  \,\F_0^\herm  \H^\herm
+ \I_{\Nc \Nr}$
is the conditional (given $\H$) covariance 
matrix of the vector $\z \in \Cset^{\Nc \Nr}$ 
collecting the signals received by all the relays, with
$\Psource>0$ and $\GlobalPrelay> 0$ denoting the 
power threshold at the source and at the destination, respectively.
The constraint on the received power at the destination
automatically limits the power expenditure at the relays.
Since problem \eqref{eq:mat-def-1} is nonconvex,
we consider its \textit{ relaxed} version:
\begin{multline}
\min_{\F_0 ,\bm F}
\Es_{\G} \left \{
\trace \left[ \left(\I_{\Nb} + \F_0^\herm  \H^{\herm}  \F^\herm  \G^\herm  \G  \,\F  \,\H  \,\F_0 \right)^{-1} \right]   \,\big |\,  \H \right\}
\\ \text{s.to}  \quad
\trace(\F_0 \, \F_0^{\herm}) \leq \Psource \quad \text{and}
\\ \quad
\Es_{\G} \left [\trace(\G  \,\F  \,\F^\herm  \G^\herm)   \,|\,  \H \right]
\leq \GlobalPrelay
\label{eq:mat-def-2}
\end{multline}
where we have used the expression of $\C$
and the inequalities
$\trace [ (\I_{\Nb}+\C^\herm \,
\Rvv^{-1} \, \C )^{-1}]
\geq \trace [ (\I_{\Nb} + \C^\herm \, \C)^{-1}]$
and $\trace(\G  \, \F \, \Rzz \, \F^\herm \G^\herm)   \le
\trace(\G  \, \F \, \F^\herm  \G^\herm) \, \trace(\Rzz)$ \cite{Horn.book.1990,Yang.2013}.
Closed-form evaluation of the cost function
in \eqref{eq:mat-def-2} is cumbersome; however, under {\bf a1)} and {\bf a2)},
it can be observed that\footnote{The proof follows easily from the facts \cite{Horn.book.1990} that
the trace of $\A$ is equal to the sum of its eigenvalues and,
if $\lambda$ is an eigenvalue of a nonsingular matrix $\A$,
then $\lambda^{-1}$ is an eigenvalue of $\A^{-1}$.
}
\begin{multline}
\trace \left[ \left(\I_{\Nb} + \F_0^\herm  \H^{\herm}  \F^\herm
\G^\herm  \G  \, \F  \, \H  \, \F_0 \right)^{-1} \right]
\\ <
\trace \left[ \left(\F_0^\herm  \H^{\herm}  \F^\herm
\G^\herm  \G  \, \F  \, \H  \, \F_0 \right)^{-1} \right]
\label{eq:traceub}
\end{multline}
where the difference between
the left- and right-hand sides tends to zero
as the minimum eigenvalue of
$\F_0^\herm  \, \H^{\herm}  \F^\herm  \G^\herm  \G  \, \F  \, \H  \, \F_0$
is significantly larger than one. This happens in
the high signal-to-noise ratio (SNR) region, i.e., when
$\Psource$ and $\GlobalPrelay$ are sufficiently large.
Relying on \eqref{eq:traceub}, we pursue a
further relaxation of \eqref{eq:mat-def-1} by replacing
$\Es_{\G} \left \{ \trace \left[ \left(\I_{\Nb} + \F_0^\herm  \H^{\herm}  \F^\herm
\G^\herm  \G  \, \F  \, \H  \, \F_0 \right)^{-1} \right]  \Big |  \H \right \} $
in \eqref{eq:mat-def-2} with  its upper bound
$\Es_{\G} \left \{ \trace \left[ \left(\F_0^\herm  \H^{\herm}  \F^\herm  \G^\herm  \G  \, \F  \, \H  \, \F_0 \right)^{-1} \right]  \Big |  \H \right \}$, which
can be evaluated
in closed-form as stated by the following Lemma.

\medskip

\begin{lemma}
\label{prop2}{
Let us assume that: {\bf a3)} $\Nd > \Nb$. Then, under {\bf a1)}, {\bf a2)}, and {\bf a3)}, it results  that
\be
\Es_{\G} \left \{
\trace \left[ \left(\F_0^\herm \H^{\herm} \F^\herm  \G^\herm  \G  \,\F  \,\H  \,\F_0 \right)^{-1} \right]  \, \big |  \, \H \right\} = \frac{\trace (\R^{-1} )}{\Nd-\Nb}
\label{eq:bound1}
\ee
where $\R \eqdef \F_0^{\herm}  \H^{\herm}  {\F}^{\herm}   {\F}  \, \H  \, \F_0
\in \Cset^{\Nb \times \Nb}$.}
\end{lemma}

\medskip

\textit{Proof:}
See Appendix~\ref{app:prop2}.
\hfill $\blacksquare$

\medskip

At this point, evaluation of the expectation in
the second constraint of
\eqref{eq:mat-def-2} is in order.
In this respect, one has
\be
\Es_{\G} \left [\trace(\F^\herm  \G^\herm  \G  \, \F)  \, |\,  \H \right]  =
\trace\left[ \Es_{\G} \left( \G^\herm  \G \right)  \F  \, \F^\herm \right]
= \trace\left(\F^\herm \F \right)
\label{eq:const}
\ee
where we have also used the cyclic property \cite{Horn.book.1990} of the trace operator.
Therefore, under  {\bf a1)}, {\bf a2)}, and
{\bf a3)}, the optimization problem \eqref{eq:mat-def-2}
can be simplified as follows
\begin{multline}
\min_{\F_0 ,\bm F} \trace \left[ \left( \F_0^{\herm}  \H^{\herm} {\F}^{\herm}  {\F}  \,\H  \, \F_0 \right)^{-1} \right]
\\
\quad \text{s.to}  \quad
\trace(\F_0 \, \F_0^{\herm}) \leq \Psource \quad \text{and} \quad
\trace\left(\F^\herm  \, \F \right) \leq \GlobalPrelay  \: .
\label{eq:mat-def-H}
\end{multline}
At this point, a comment regarding the constraints in
\eqref{eq:mat-def-1} and  \eqref{eq:mat-def-H}
is in order. The constraint $\trace(\F_0 \, \F_0^{\herm}) \leq \Psource$
in  \eqref{eq:mat-def-1} and  \eqref{eq:mat-def-H}
limits the average transmitted
power of the source and it is standard in the design of linear MIMO
transceivers \cite{Palomar.2003}. Regarding
the second constraint in \eqref{eq:mat-def-1}, we observe that,
given $\H$ and $\G$,
$\mathcal{P}(\H,\G) \eqdef \trace(\G  \,\F  \,\Rzz  \,\F^\herm  \G^\herm)$ represents
the average received power at the destination. It is noteworthy
that  $\mathcal{P}(\H,\G)$ is typically limited in
those scenarios
where a target performance has to be achieved and
per-node fairness is not of concern \cite{Goldsmith,Merched.2008}.
The constraint $\trace\left(\F^\herm  \, \F \right) \leq
\GlobalPrelay$ in \eqref{eq:mat-def-H}, which has been obtained by averaging
a relaxed version of $\mathcal{P}(\H,\G)$ with respect to the probability
distribution of $\G$, fixes a limit on the total average power transmitted by the
relays, so-called \emph{sum-power constraint} \cite{Champagne2013}.\footnote{Design with per-relay power constraints can be solved by properly reformulating the
problem into an  equivalent  optimization with  a  sum-power  constraint
\cite{Sartenaer,Cheng}.}

To solve \eqref{eq:mat-def-H}, we use
the following Lemma.

\medskip

\begin{lemma}\label{prop3}{
For a positive definite matrix $\A \in \Cset^{n \times n}$, 
the following inequality holds:
\be
\trace(\A^{-1}) \ge \sum_{\ell=1}^m \frac{1}{\{\A\}_{\ell\ell}}
\ee
where $\{\A\}_{\ell\ell}$ is the $\ell$th diagonal entry of $\A$ and the inequality
is achieved if $\A$ is diagonal.}
\end{lemma}

\medskip

\textit{Proof:}
See \cite[p. 65]{Kay}.
\hfill $\blacksquare$

\medskip

As a consequence of Lemma~\ref{prop3},
the minimum value of the cost function in \eqref{eq:mat-def-H}
is achieved if $\F_0^\herm \,  \A \, \F_0$ is
diagonal, with $\A \eqdef \H^{\herm} {\F}^{\herm} {\F} \, \H \in \Cset^{\Ns \times \Ns}$.
In what follows, we consider
three different approaches to achieve the desired 
diagonalization of $\F_0^\herm \,  \A \, \F_0$:
the first one is based on the SVD of the
composite matrix $\H =[\H_1^{ \trasp}, \H_2^{\trasp}, \ldots,
\H_{\Nc}^{ \trasp}]^{\trasp}$ and it results in a (possible) selection
of all the relays; 
the second one relies on the SVDs of the
individual matrices $\H_1, \H_2, \ldots, \H_{\Nc}$, thus leading to
a single-relay selection; 
the last one exploits the
SVDs of row-based partitions of $\H$
and it can be interpreted as a joint antenna-and-relay selection scheme.

\subsection{Design based on the SVD of the composite first-hop channel matrix}
\label{sec:SVD-H}

One can attempt to recruit all the relays in the second hop
of the cooperative scheme
by diagonalizing $\F_0^\herm \,  \A \, \F_0$ through
the SVD $\H = \bm{U}_{\text{h}} \,
[\bm{O}_{\Ns \times (\Nc \Nr -\Ns)}, \bm{\Lambda}_{\text{h}}]^\trasp \, \bm{V}_{\text{h}}^\herm$ of $\H$, where the matrices
$\bm{U}_{\text{h}} \in \Cset^{(\Nc \Nr) \times (\Nc \Nr)}$
and $\bm{V}_{\text{h}} \in \Cset^{\Ns \times \Ns}$ are unitary,
and
$\bm{\Lambda}_{\text{h}} \eqdef \diag[\lambda_{{\text{h}}}(1), \lambda_{\text{h}}(2), \ldots, \lambda_{\text{h}}(\Ns)]$ gathers
the corresponding nonzero singular values arranged in increasing order.
By substituting the SVD of $\H$ in $\A$, it follows by direct inspection
that $\F_0^\herm \,  \A \, \F_0$
is diagonal if (see, e.g., \cite{Israel.book.2002})
\barr
\F_0 & = \V_{\text{h},\text{right}} \, \bm{\Omega}^{1/2}
\label{eq:F0male}
\\
\F_i & = \Q_i \, \bm{\Delta}_i^{1/2} \, \U_{\text{h},\text{right},i}^\dag
\label{eq:Fimale}
\earr
where $\V_{\text{h},\text{right}} \in \Cset^{\Ns \times \Nb}$
contains the $\Nb$ rightmost columns from $\bm{V}_{\text{h}}$, the matrices
$\bm{\Omega} \eqdef \diag[\omega(1), \omega(2), \ldots, \omega(\Nb)]$
and $\bm{\Delta}_i \eqdef \diag[\delta_i(1), \delta_i(2), \ldots, \delta_i(\Ns)]$
are determined in a second step, for $i \in \{1,2,\ldots, \Nc\}$, the arbitrary matrix
${\Q}_{i} \in \Cset^{\Nr \times \Ns}$
obeys $\bm{Q}_{i}^\herm \bm{Q}_{i} =\I_{\Ns}$,  provided that $\Ns \le \Nr$,
$\U_{\text{h},\text{right}} \eqdef [\U_{\text{h},\text{right},1}^\trasp,
\U_{\text{h},\text{right},2}^\trasp, \ldots, \U_{\text{h},\text{right},\Nc}^\trasp]^\trasp
\in \Cset^{(\Nc \Nr) \times \Ns}$ contains the $\Ns$ rightmost columns from 
$\bm{U}_{\text{h}}$, with the matrix $\U_{\text{h},\text{right},i} \in \Cset^{\Nr \times \Ns}$ being full-column rank.

Using \eqref{eq:F0male} and \eqref{eq:Fimale}, problem
\eqref{eq:mat-def-H} ends up to
\begin{multline}
\min_{ \bm{\omega}, \{\bm{\delta}_i\}_{i=1}^\Nc}
f_0 \left(\bm{\omega},\{\bm{\delta}_i\}_{i=1}^\Nc\right) \quad
\text{s.to} \quad
\sum_{\ell=1}^{\Nb} \omega(\ell) \leq \Psource, \,
\omega(\ell) > 0, \\ \text{and} \quad
\sum_{i=1}^{\Nc} \sum_{\ell=1}^{\Ns} \delta_{i}(\ell)
\left(\U_{\text{h},\text{right},i}^\herm \, \U_{\text{h},\text{right},i} \right)^{-1}_{\ell \ell} \leq \GlobalPrelay, \,
\delta_i(\ell) > 0
\label{eq:opt-0}
\end{multline}
where we have defined $\bm{\omega} \eqdef [\omega(1), \omega(2), \ldots, \omega(\Nb)]^T \in \Rset^{\Nb}$,
$\bm{\delta}_i \eqdef [\delta_i(1), \delta_i(2), \ldots, \delta_i(\Ns)] \in \Rset^{\Ns}$,
for $i \in \{1,2,\ldots, \Nc\}$,
\be
f_0 \left(\bm{\omega},\{\bm{\delta}_i\}_{i=1}^\Nc\right) \eqdef  \sum_{\ell=1}^{\Nb} \frac{1}{\omega(\ell)
\, \lambda_{{\text{h}}}^2(\Delta N+\ell) \displaystyle\sum_{i=1}^{\Nc}
\delta_i(\Delta N+\ell)}
\label{eq:costmale}
\ee
with $\Delta N \eqdef \Ns-\Nb \ge 0$.
All the inequality constraints in \eqref{eq:opt-0} are linear.
However, it is shown in Appendix~\ref{app:2} that, when $\Nc >1$, the cost function \eqref{eq:costmale} is the sum of
$\Nb$ functions that are neither strictly convex
nor strictly concave on $\Rset_{+}^{n+1}$.
Hence, trying to solve \eqref{eq:opt-0} with the available
optimization tools leads to poor performance 
in multiple-relay WSNs.


\subsection{Design based on the SVD of the 
individual first-hop channel matrices}
\label{sec:SVD-Hi}

A simple design can be developed by setting
$\F_{i} =\bm{O}_{\Nr \times \Nr}$,
for each $i \in \{1,2,\ldots, \Nc\}-\istar$.
Basically, such a choice leads to
a \textit{ single-relay selection} scheme \cite{Darsena.2017}, which imposes
that only one relay (i.e., that for $i = \istar$)
is recruited to transmit and all the remaining ones
keep silent in the second hop.

Herein, we assume that
$\H_{i}$ is full-column rank, i.e.,
$\rank(\H_{i})=\Ns \le \Nr$,
for each $i \in \{1,2,\ldots, \Nc\}$.
Let
\be
\U_{\text{h},i} \, [\bm{O}_{\Ns \times (\Nr-\Ns)}, \bm{\Lambda}_{\text{h},i}]^\trasp
\, \V_{\text{h},i}^\herm 
\ee
be the SVD of $\H_{i}$,
where
\be
\bm{\Lambda}_{\text{h},i}
\eqdef \diag[\lambda_{{\text{h},i}}(1), \lambda_{\text{h},i}(2), \ldots, \lambda_{\text{h},i}(\Ns)]
\ee
contains the singular values of $\H_{i}$, arranged in increasing
order,  and the unitary matrices $\U_{\text{h},i} \in \Cset^{\Nr \times \Nr}$
and $\V_{\text{h},i} \in \Cset^{\Ns \times \Ns}$
collect the corresponding left and right singular vectors, respectively.
In this case, one has
$\A=\H_{\istar}^{\herm} \F_{\istar}^{\herm} \F_{\istar} \, \H_{\istar}$
and, by substituting the SVD of $\H_{\istar}$ in this matrix equation, one has
that the diagonalization of $\F_0^\herm \,  \A \, \F_0$ is ensured by
\barr
\F_0 & = \V_{\text{h}, \istar, \text{right}} \, \bm{\Omega}^{1/2}
\label{eq:solF0-single}
\\
\F_{\istar} &  = {\Q}_{\istar} \bm{\Delta}^{1/2} \, \U_{\text{h},\istar,\text{right}}^\herm
\label{eq:solFi-single}
\earr
where
$\U_{\text{h}, \istar, \text{right}} \in \Cset^{\Nr \times \Ns}$
and
$\V_{\text{h}, \istar, \text{right}} \in \Cset^{\Ns \times \Nb}$
contain the $\Ns$ and $\Nb$ rightmost columns from $\U_{\text{h}, \istar}$
and $\V_{\text{h}, \istar}$, respectively, ${\Q}_{\istar} \in \Cset^{\Nr \times \Ns}$
is an arbitrary matrix obeying
$\bm{Q}_{\istar}^\herm \bm{Q}_{\istar} =\I_{\Ns}$,
$\bm{\Omega}$ has been defined in Subsection~\ref{sec:SVD-H},
and
$\bm{\Delta} \eqdef \diag[\delta(1), \delta(2), \ldots, \delta(\Ns)]$.
To fully specify the solution of \eqref{eq:mat-def-H}
in the case of single-relay selection,
optimization of $\bm{\Omega}$, $\bm{\Delta}$, and $\istar$
is accomplished in two steps.

First, for a given $\istar \in \{1,2,\ldots, \Nc\}$,
by substituting \eqref{eq:solF0-single} and \eqref{eq:solFi-single} in
\eqref{eq:mat-def-H}, one obtains the
scalar optimization problem with linear inequality constraints:
\begin{multline}
\min_{ \bm{\omega}, \bm{\delta}}
f_1\left(\istar, \bm{\omega}, \bm{\delta} \right) \quad
\text{s.to} \quad
\sum_{\ell=1}^{\Nb} \omega(\ell) \leq \Psource, \:
\omega(\ell) > 0, \\ \text{and} \quad
\sum_{\ell=1}^{\Ns} \delta(\ell) \leq \GlobalPrelay, \:
\delta(\ell) > 0
\label{eq:opt-1}
\end{multline}
with
\be
f_1\left(\istar, \bm{\omega}, \bm{\delta} \right)
\eqdef \sum_{\ell=\Ns-\Nb+1}^{\Ns}
\frac{1}{\lambda_{\text{h}, \istar}^2(\ell) \,
\omega(\ell)  \, \delta(\ell)}
\ee
where
$\bm{\omega}$ has been previously defined in Subsection~\ref{sec:SVD-H}
and  $\bm{\delta} \eqdef [\delta(1), \delta(2),
\ldots, \delta(\Ns)]^T \in \Rset^{\Ns}$.
Since $f_1\left(\istar, \bm{\omega}, \bm{\delta} \right)$ is a convex
function (see Appendix~\ref{app:2}), the
optimization problem \eqref{eq:opt-1} is convex
and, thus, its solution $\bm{\omega}_\text{opt}(\istar)$ and
$\bm{\delta}_\text{opt}(\istar)$
can be found  by using efficient numerical techniques \cite{Boyd.book.2004}.
For instance, if one resorts to interior point methods,
convergence  arbitrarily close to the optimal solution is achieved 
in a number of iterations that is proportional to the logarithm 
of the problem dimension \cite{Colombo.2008},
with a complexity per iteration dictated by the cost
$M$ of computing a Newton direction \cite{Roger.1987}.

Second, the optimal value $i_\text{opt}$ of $\istar$  is obtained as
\be
i_\text{opt} \eqdef \arg\min_{\istar \in \{1,2,\ldots,\Nc\}}
f_1\left(\istar, \bm{\omega}_\text{opt}(\istar), \bm{\delta}_\text{opt}(\istar) \right)
\label{eq:selection-1}
\ee
which allows one to single out the best relay
among the $\Nc$ available ones.
The solution of \eqref{eq:selection-1}
can be obtained by solving \eqref{eq:opt-1}
for each $\istar \in \{1,2,\ldots, \Nc\}$, with an overall complexity
$\mathcal{O}\left[\Nc \,M \log(\Nb+\Ns)\right]$.

In the SISO configuration, i.e., when $\Nb=\Ns=\Nr=\Nd=1$, and when there is no precoding
at the source, i.e., $\F_0 \equiv f_0=\sqrt{\Psource}$, one gets
$\F_{\istar}=\diag(0, \ldots, 0, f_{\istar}, 0, \ldots, 0)$ and \eqref{eq:selection-1} boils down to
$i_\text{opt} \eqdef \arg\max_{\istar \in \{1,2,\ldots,\Nc\}}
\{|h_i|^2\}$, with $h_i$ denoting the channel coefficients between the
source and the $i$th relay. According to \cite{Ribeiro}, such a scheme
has a full diversity order equal to $\Nc$.
However, as we will see in Section~\ref{sec:simulation}, such a design suffers from
a diversity loss in a MIMO setting, i.e., when $\Ns, \Nr, \Nd >1$.

\subsection{Design based on the SVD of row-based partitions of
the composite first-hop channel matrix}
\label{sec:SVD-Hpart}

In the considered cooperative MIMO WSN, there are $\Nc$ relays equipped with $\Nr$ antennas,
which amounts to a total number of $\Nc \, \Nr$ distributed antennas. Here,
we propose to choose the best $\Nb=\Ns$ antennas out of the $\Nc \, \Nr$ ones.\footnote{Our
design can be simply extended to the case $\Ns \ge \Nb$.}
Such antennas can either be physically located on a single relay, or
be spatially distributed over different relays, 
thus accomplishing a joint
antenna-and-relay selection scheme.

Let $\mathcal{S} \eqdef \left\{(n,i), \forall n \in \{1,2,\ldots, \Nr\},
\forall i \in \{1,2,\ldots,\Nc\}\right\}$ collect all
the $\Nc \, \Nr$ antenna elements in the network, with
the generic (ordered) pair $(n,i)$ uniquely
identifying the $n$th antenna located on the
$i$th relay.
The number of distinct subsets of $\mathcal{S}$
that have exactly $\Nb$ elements is given by
the binomial  coefficient
$Q \eqdef \binom{\Nc \Nr}{\Nb}$.\footnote{The empty set $\emptyset$ and the set
$\mathcal{S}$ are considered as subsets of $\mathcal{S}$ as well.}
By excluding the trivial choice $\emptyset$ and
the degenerate case $\mathcal{S}$ (discussed in Subsection~\ref{sec:SVD-H}) ,
we denote with
\be
\mathcal{S}^{(q)} \eqdef \left\{(n_1^{(q)},i_1^{(q)}), (n_2^{(q)},i_2^{(q)}),
\ldots, (n_\Nb^{(q)},i_\Nb^{(q)})\right\}
\ee
the selected subset of $\mathcal{S}$, with $q \in \{1,2,\ldots, Q-2\}$,
obeying $\mathcal{S}^{(q_1)} \neq \mathcal{S}^{(q_2)}$ for
$q_1 \neq q_2$.
Additionally, we use the notation $N_i^{(q)} \in \{0,1, \ldots, \Nb\}$ to indicate the
number of pairs of $\mathcal{S}^{(q)}$ having the same second entry: in other words,
$N_i^{(q)}$ represents the number
of antennas activated on the $i$th relay according to the $q$th selection.
It results that
$\sum_{i=1}^\Nc N_i^{(q)} = \Nb$.

The selected antennas generate a first-hop channel matrix
\be
\H^{(q)} \eqdef
[(\H_1^{(q)})^{ \trasp}, (\H_2^{(q)})^{ \trasp}, \ldots,
(\H_\Nc^{(q)})^{ \trasp}]^{\trasp}
\in \Cset^{\Nb \times \Nb} 
\ee
and
a relaying matrix
\be
\F^{(q)} \eqdef \diag(\F_1^{(q)}, \F_2^{(q)}, \ldots, \F_{\Nc}^{(q)})
\in \Cset^{\Nb \times \Nb} 
\ee
with
$\H_i^{(q)} \in \Cset^{N_i^{(q)} \times \Nb}$
and $\F_i^{(q)} \in \Cset^{N_i^{(q)} \times N_i^{(q)}}$.
By convention, if $N_i^{(q)}=0$, then $\H_i^{(q)}$
and $\F_i^{(q)}$ are empty matrices.

With reference to the $q$th selection,
we formulate a new optimization problem,
for $q \in \{1,2,\ldots, Q-2\}$,
which is formally obtained
from \eqref{eq:mat-def-H} by replacing $\H$ and $\F$ with
$\H^{(q)}$ and $\F^{(q)}$, respectively, whose cost function
achieves its minimum  value if $\F_0^\herm \,  \A^{(q)} \, \F_0$
is diagonal (see Lemma~\ref{prop3}), with
$\A^{(q)} \eqdef  (\H^{(q)})^{\herm} \, (\F^{(q)})^{\herm}
\, \F^{(q)} \, \H^{(q)}$.
For $i \in \{1,2,\ldots, \Nc\}$,
let $\H_i^{(q)} = \bm{U}_{\text{h},i}^{(q)} \,
[\bm{O}_{N_i^{(q)} \times (\Nb -N_i^{(q)})}, \bm{\Lambda}_{\text{h},i}^{(q)}] \,
(\bm{V}_{\text{h},i}^{(q)})^\herm$ be the SVD of the (nonempty) matrix $\H_i^{(q)}$,
which is assumed to be full-row rank, i.e., $\rank(\H_i^{(q)})=N_i^{(q)}$,
where $\bm{U}_{\text{h},i}^{(q)} \in \Cset^{N_i^{(q)} \times N_i^{(q)}}$
and $\bm{V}_{\text{h},i}^{(q)} \in \Cset^{\Nb \times \Nb}$ are unitary,
and  the diagonal matrix
$\bm{\Lambda}_{\text{h},i}^{(q)} \eqdef \diag[\lambda_{{\text{h},i}}^{(q)}(1), \lambda_{\text{h},i}^{(q)}(2), \ldots, \lambda_{\text{h},i}^{(q)}(N_i^{(q)})]$
collects the corresponding nonzero singular values arranged in increasing order.
In this case, the diagonalization of $\F_0^\herm \,  \A^{(q)} \, \F_0$ can be obtained
by resorting to the following structures
\barr
\F_0 & = [(\V_{\text{h},\text{right}}^{(q)})^\herm]^{-1} \, \bm{\Omega}^{1/2}
\label{eq:F0bene}
\\
\F_i^{(q)} & = \Q_i \, \bm{\Delta}_i^{1/2} \, (\U_{\text{h},i}^{(q)})^\herm
\label{eq:Fibene}
\earr
where
\be
\V_{\text{h},\text{right}}^{(q)} \eqdef [\V_{\text{h},\text{right},1}^{(q)},
\V_{\text{h},\text{right},2}^{(q)}, \ldots,
\V_{\text{h},\text{right},\Nc}^{(q)}]
\in \Cset^{\Nb \times \Nb}
\ee
with
$\V_{\text{h},\text{right},i}^{(q)} \in \Cset^{\Nb \times N_i^{(q)}}$ gathering
the $N_i^{(q)}$ rightmost columns from $\bm{V}_{\text{h},i}^{(q)}$,
${\Q}_{i} \in \Cset^{N_i^{(q)} \times N_i^{(q)}}$ is an arbitrary unitary matrix,
$\bm{\Omega}$ and $\bm{\Delta}_i$ have been defined in Subsection~\ref{sec:SVD-H}.

To optimize $\bm{\Omega}$, $\bm{\Delta}_i$, and
$q$, we resort to a two-step procedure as in the previous
subsection.
By substituting \eqref{eq:F0bene}--\eqref{eq:Fibene} in
\eqref{eq:mat-def-H} (with $\H^{(q)}$ and $\F^{(q)}$ in lieu of
$\H$ and $\F$, respectively),
for a given $q \in \{1,2,\ldots,Q-2\}$,
one gets the convex
optimization problem (see Appendix~\ref{app:2}) with linear inequality constraints:
\begin{multline}
\min_{ \bm{\omega}, \{\bm{\delta}_i\}_{i=1}^\Nc}
f_2\left(q, \bm{\omega}, \{\bm{\delta}_i\}_{i=1}^\Nc\right) \quad
\text{s.to}
\\
\sum_{\ell=1}^{\Nb} \omega(\ell) \left[(\V_{\text{h},\text{right}}^{(q)})^\herm \,
\V_{\text{h},\text{right}}^{(q)}\right]^{-1}_{\ell \ell} \leq \Psource, \:
\omega(\ell) > 0, \\ \text{and} \quad
\sum_{i=1}^{\Nc} \sum_{\ell=1}^{\Ns} \delta_{i}(\ell)   \leq \GlobalPrelay, \,
\delta_i(\ell) > 0
\label{eq:opt-2}
\end{multline}
with
\be
f_2\left(q, \bm{\omega}, \{\bm{\delta}_i\}_{i=1}^\Nc \right)
\eqdef \sum_{i=1}^{\Nc}
\sum_{\ell=1}^{\Nb}
\frac{1}{\left[\lambda_{{\text{h},i}}^{(q)}(\ell)\right]^2
 \omega_i(\ell) \, \delta_i(\ell)}
\ee
where $\omega_i(\ell) \eqdef \omega\left(\sum_{m=1}^{i-1} N_m^{(q)}+\ell\right)$,
whereas $\bm{\omega}$ and $\bm{\delta}_i$ have been
defined in Subsection~\ref{sec:SVD-H}.
Similarly to problem \eqref{eq:opt-1}, the
solution $\bm{\omega}_\text{opt}(q)$ and
$\{\bm{\delta}_{i,\text{opt}}(q)\}_{i=1}^\Nc$ of \eqref{eq:opt-2}
can be found  by using, e.g., interior point methods \cite{Boyd.book.2004}.
Finally, the best value $q_\text{opt}$ of $q$ is found by solving
\be
q_\text{opt} \eqdef \arg\min_{q
\in \{1,2,\ldots,Q-2\}}
f_2\left(q, \bm{\omega}_\text{opt}(q),
\{\bm{\delta}_{i,\text{opt}}(q)\}_{i=1}^\Nc \right)
\label{eq:selection-2}
\ee
which determines the best $\Nb$-dimensional subset of the available
$\Nc \, \Nr$ antennas.
The solution of \eqref{eq:selection-2}
can be obtained by solving \eqref{eq:opt-2}
for each $q \in \{1,2,\ldots,Q-2\}$, with an overall complexity
$\mathcal{O}\left[(Q-2) \,M \log(\Nb+\Nb\,\Nc)\right]$, which
is larger than that required to select the best relay
(see Subsection~\ref{sec:SVD-Hi}), especially for large number of relays.

When $\Nb=\Ns=\Nr=1$, the optimization
problems \eqref{eq:opt-1}-\eqref{eq:selection-1} and
\eqref{eq:opt-2}-\eqref{eq:selection-2} yield the same solution
and, thus, the design \eqref{eq:opt-2}-\eqref{eq:selection-2} exhibits
full diversity order $\Nc$, too.
However, we will show in the next section that,
when $\Nb, \Ns, \Nr >1$ (MIMO WSN), the proposed joint
antenna-and-relay selection scheme ensures a significant
performance improvement with respect to single-relay selection,
in terms of both diversity order and coding gain.

\begin{figure}
\centering
\includegraphics[width=\linewidth, trim=20 20 20 20]{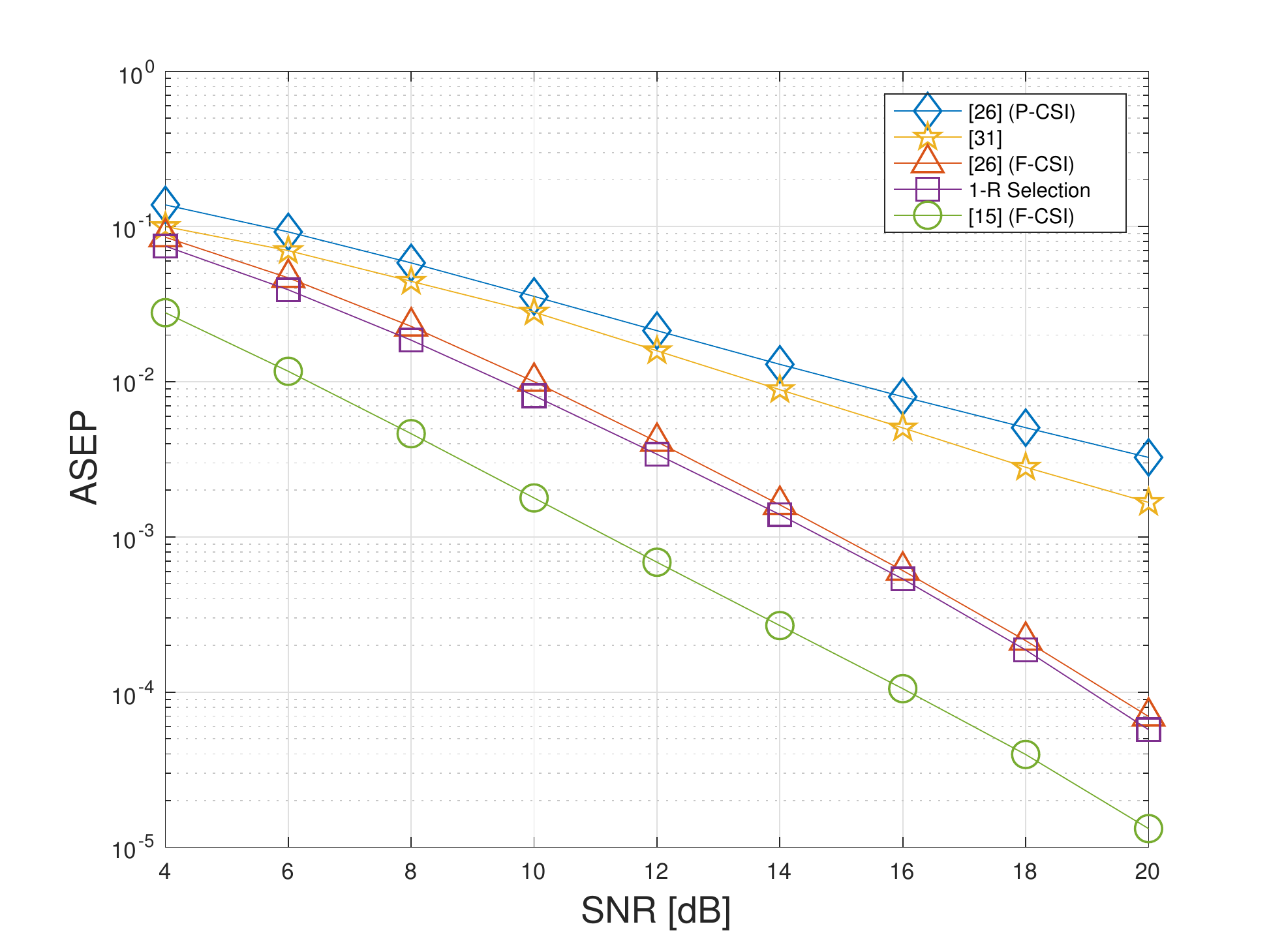}
\caption{ASEP versus $\text{SNR}$ (Example~1: $N=1$ and $\Nc=2$).}
\label{fig:fig_1}
\end{figure}
\begin{figure}
\centering
\includegraphics[width=\linewidth, trim=20 20 20 20]{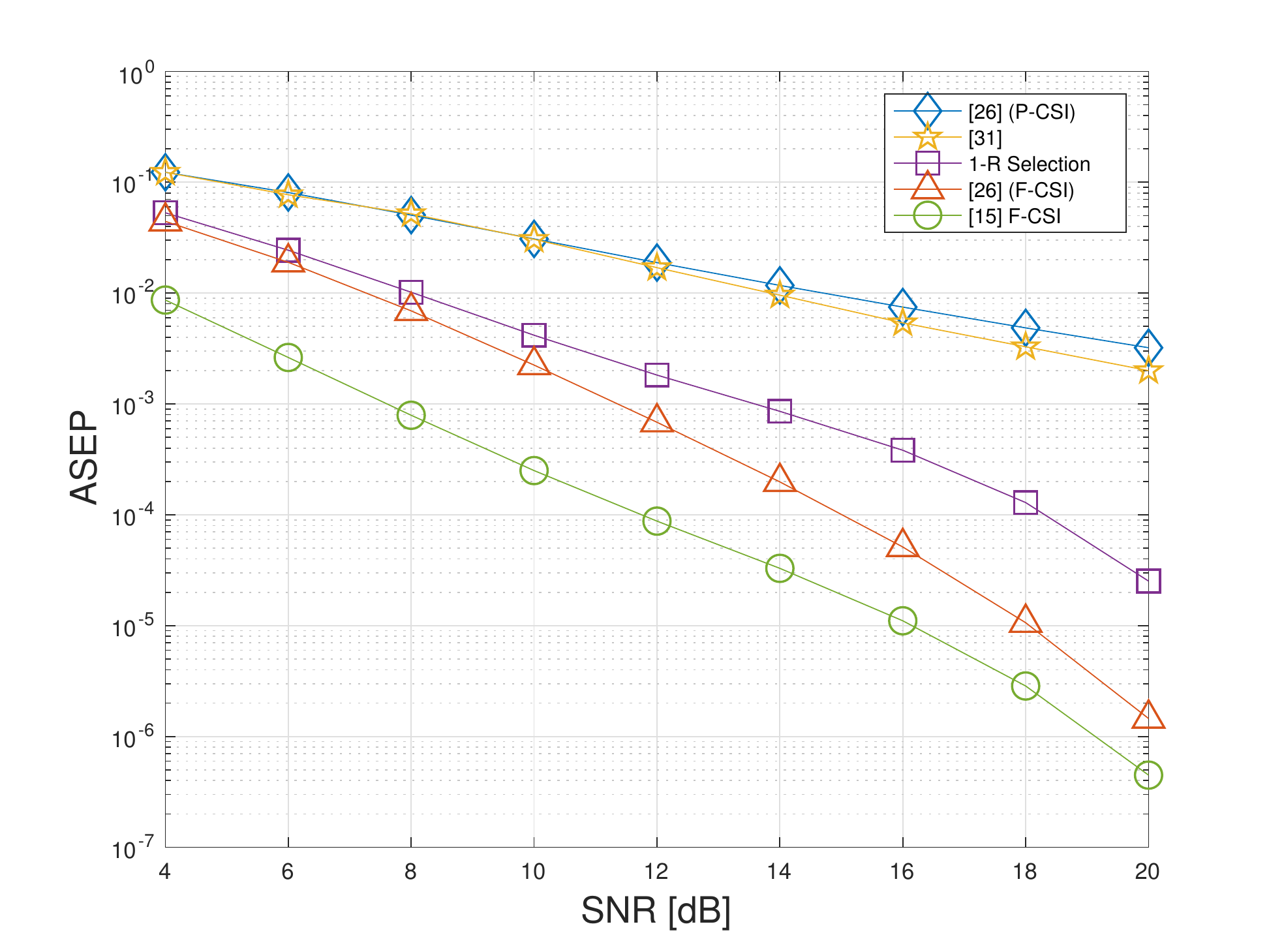}
\caption{ASEP versus $\text{SNR}$  (Example~1: $N=1$ and $\Nc=3$).}
\label{fig:fig_2}
\end{figure}

\section{Numerical results}
\label{sec:simulation}

In this section, to assess the performance of the considered P-CSI designs, we present
the results of Monte Carlo computer
simulations, aimed at evaluating the ASEP of the corresponding cooperative systems,
transmitting quadrature phase-shift-keying (QPSK) symbols.
We set $N \eqdef \Nb=\Ns=\Nr=\Nd$ in all the forthcoming examples,
with $N \in \{1,2,3\}$.
We also assume that $\Psource=\GlobalPrelayup =\Ptot$.
Consequently,  the SNR is defined as
$\text{SNR} \eqdef \Ptot$, which measures the per-antenna
link quality of both the first- and second-hop transmissions.
Besides the single-relay selection method described in
Subsection~\ref{sec:SVD-Hi},
referred to as ``$1$-R Selection", and the
joint
antenna-and-relay selection scheme 
developed in Subsection~\ref{sec:SVD-Hpart},
referred to as ``JAR Selection", we also report 
the performance of
\cite[CSI Assumptions I and II]{Yi.2007} in the case of single-antenna nodes
(i.e., $N=1$)
and that of \cite{Danaee2012} for both single- and multiple-antennas nodes
(i.e., $N \in \{2,3\}$).
As a reference lower bound, we additionally include in all the plots
the ASEP curves of  the F-CSI design proposed in \cite{Darsena-jcn},
whose design relies on the
additional knowledge of the $i$th
second-hop channel matrix $\G_i$ at the $i$th relay,
for $i \in \{1,2,\ldots, \Nc\}$. This F-CSI method
exhibits a theoretical diversity order equal to $\Nc \, \Nr -\Nb +1$
\cite{Darsena-jcn}.

The ASEP has been evaluated by carrying out $10^3$ independent Monte Carlo trials,
with each run using independent sets of channel realizations and noise, and an
independent record of $10^6$ source symbols.

\subsection{Example~1: Single-antenna nodes}

We report in Figs.~\ref{fig:fig_1} and
\ref{fig:fig_2} the ASEP performance of the considered
schemes as a function of the SNR, for single-antenna nodes
(i.e., $N=1$) and two different values of the number of
relays $\Nc \in \{2,3\}$.
We would like to remember that, in the case of $N=1$, the
two approaches ``$1$-R Selection" and ``JAR Selection"
are equivalent and, thus, only the performance of the
``$1$-R Selection" method are reported.

Results clearly show that no diversity is achieved by
\cite{Yi.2007} (CSI Assumption II
corresponding to P-CSI) and \cite{Danaee2012},
irrespective of the number of relays. 
On the other hand, the ``$1$-R Selection" scheme exhibits the same
diversity order of the F-CSI methods proposed in
\cite{Yi.2007} (CSI Assumption I) and \cite{Darsena-jcn},
which linearly increases with $\Nc$.
This fact allows the ``$1$-R Selection" design to significantly
outperform both \cite{Yi.2007} (P-CSI) and \cite{Danaee2012}, which
rely on the same amount of CSI.
Remarkably, the ``$1$-R Selection" scheme performs comparably to
\cite{Yi.2007} (F-CSI) in the case of $\Nc=2$ relays.
Compared to single-relay selection,
the performance
improvement of the F-CSI approaches -- arising
from the additional instantaneous knowledge of the second-hop matrix $\G$ --
becomes more and more apparent when the number of relays $\Nc$ increases.

\begin{figure}
\centering
\includegraphics[width=\linewidth, trim=20 20 20 20]{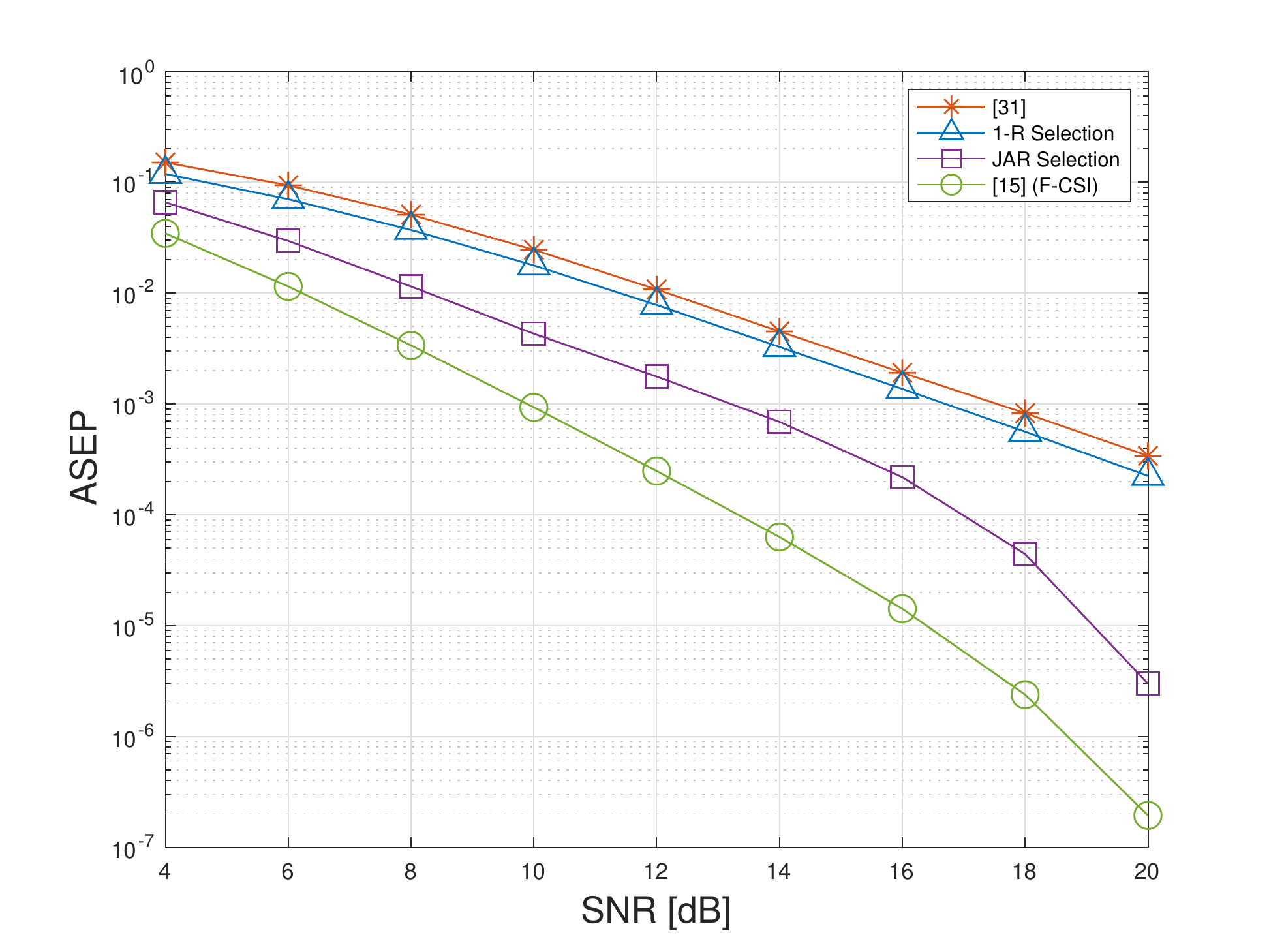}
\caption{ASEP versus $\text{SNR}$ (Example~2: $N=2$ and $\Nc=2$).}
\label{fig:fig_3}
\end{figure}
\begin{figure}
\centering
\includegraphics[width=\linewidth, trim=20 20 20 20]{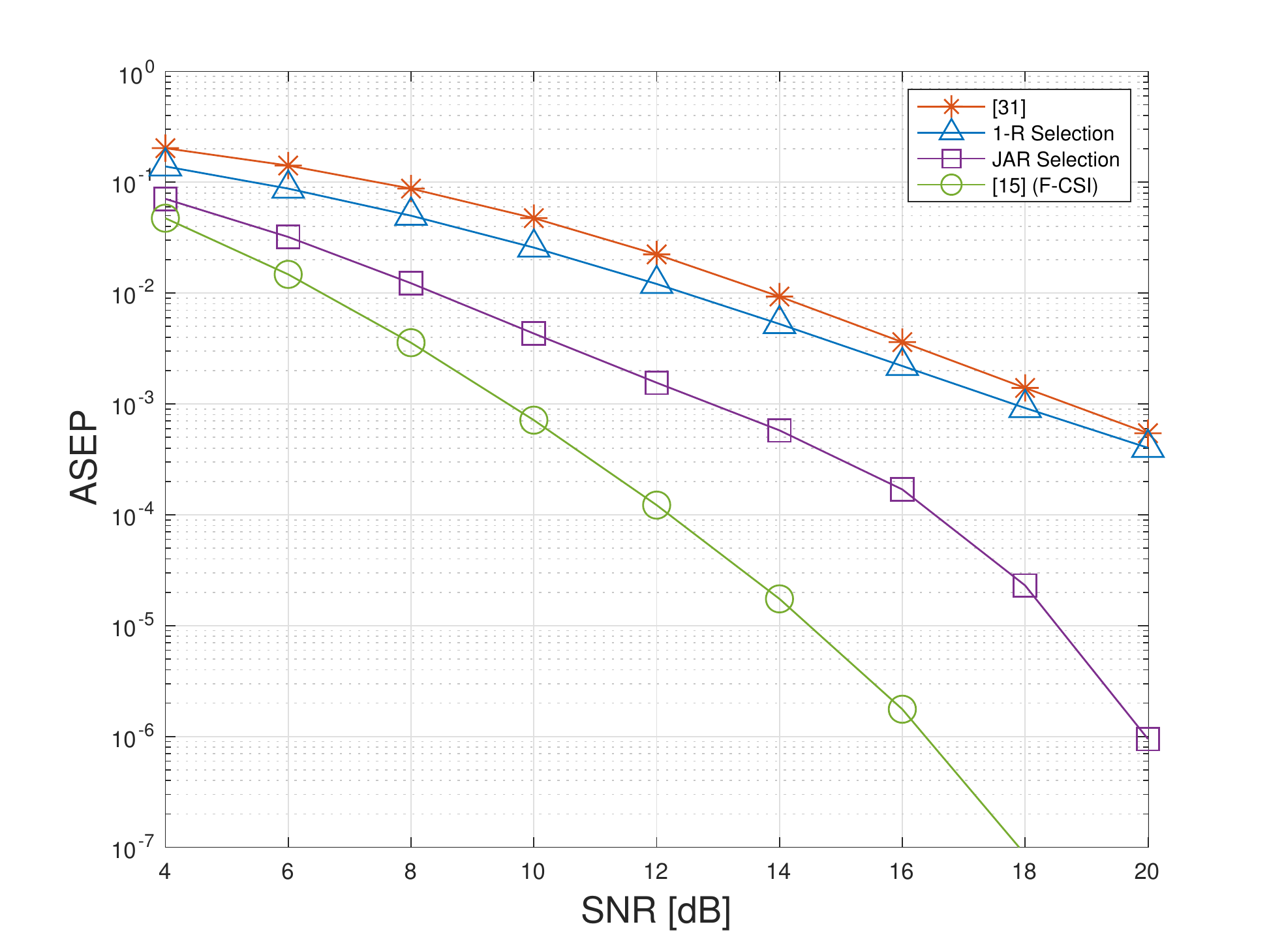}
\caption{ASEP versus $\text{SNR}$  (Example~2: $N=2$ and $\Nc=3$).}
\label{fig:fig_4}
\end{figure}
\begin{figure}
\centering
\includegraphics[width=\linewidth, trim=20 20 20 20]{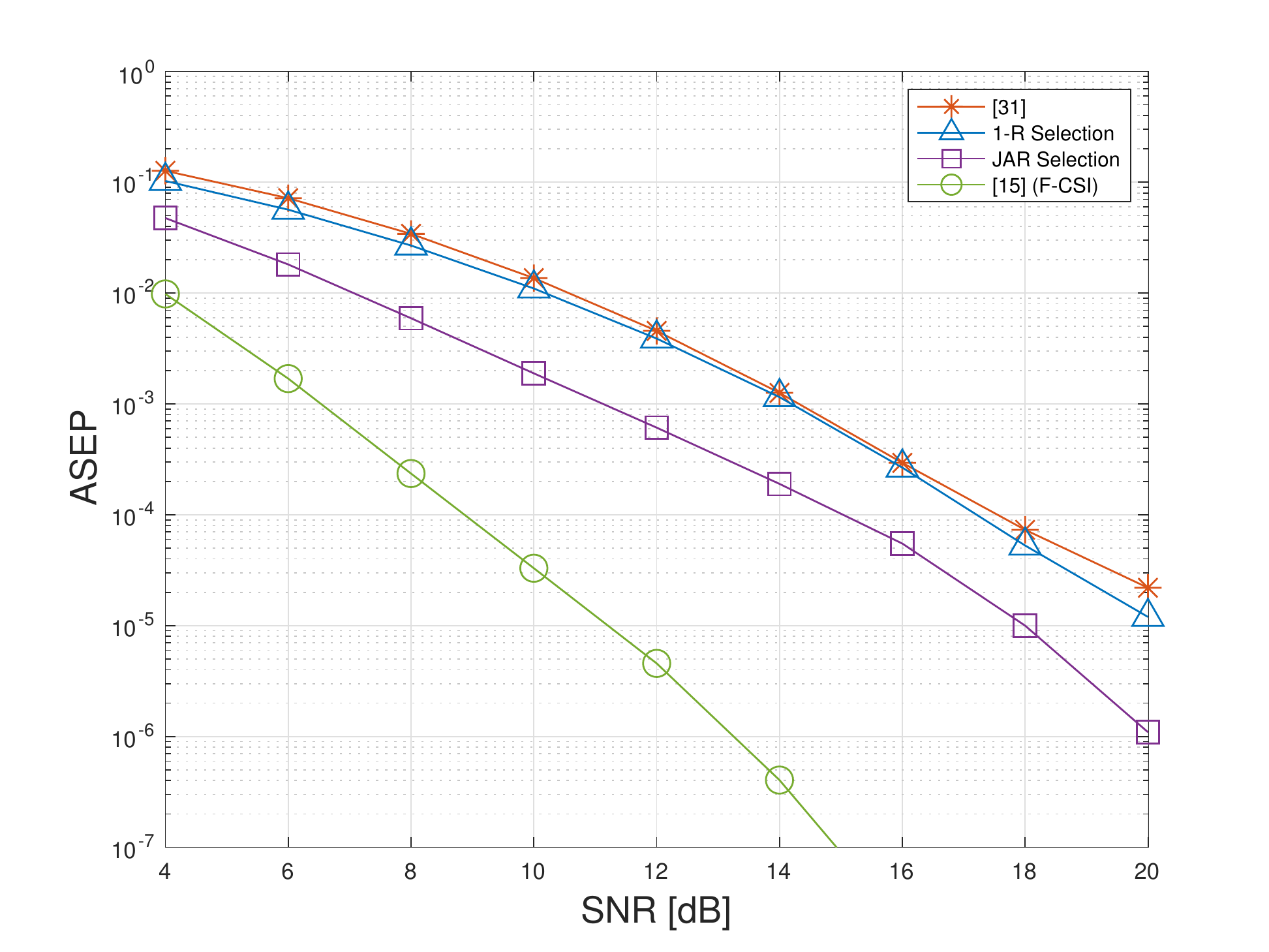}
\caption{ASEP versus $\text{SNR}$ (Example~2: $N=3$ and $\Nc=2$).}
\label{fig:fig_5}
\end{figure}
\begin{figure}
\centering
\includegraphics[width=\linewidth, trim=20 20 20 20]{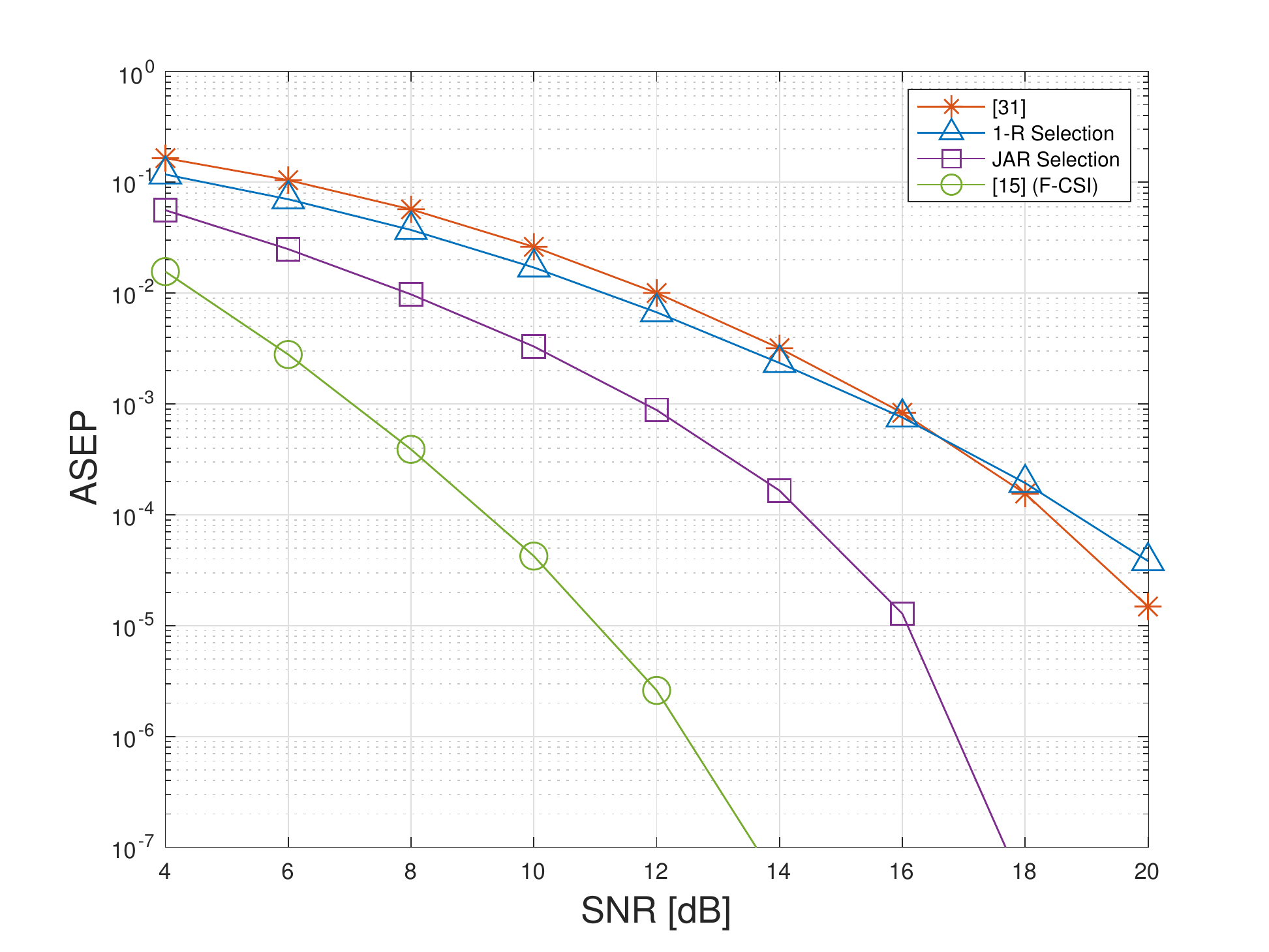}
\caption{ASEP versus $\text{SNR}$  (Example~2: $N=3$ and $\Nc=3$).}
\label{fig:fig_6}
\end{figure}

\subsection{Example~2: Multiple-antenna nodes}

Figs.~\ref{fig:fig_3}, \ref{fig:fig_4},
\ref{fig:fig_5}, and \ref{fig:fig_6}
show the ASEP performance of the
considered designs as a function of the SNR, for
two different multi-antenna configurations $N \in \{2,3\}$
and two different values of the number of
relays $\Nc \in \{2,3\}$, respectively.

It is apparent from these plots that,
in a multi-antenna deployment,
the ``$1$-R Selection" approach and  \cite{Danaee2012}
perform comparably, both exhibiting a diversity
loss with respect to the F-CSI design \cite{Darsena-jcn}.
As claimed, especially  in the high SNR regime,
the proposed  ``JAR Selection" design significantly outperforms both
the ``$1$-R Selection" scheme and  \cite{Danaee2012}, under the
same amount of P-CSI. Such a performance gap
remarkably scales up as the number of antennas at the nodes
increases from $N=2$ to $N=3$.
Interestingly, the diversity order of the ``JAR Selection" scheme
increases with $\Nc$, as in 
\cite{Darsena-jcn} which, however, requires F-CSI.

\section{Conclusions}
\label{sec:concl}

We studied the problem of designing multi-relay AF cooperative WSNs,
based on the knowledge of the instantaneous
values of the first-hop MIMO channel matrix and
statistical characterization of the second-hop one
(partial CSI scenario).
In this case, antenna/relay selection schemes arise necessarily to
formulate mathematically tractable 
design problems, which can be solved by using
standard convex optimization tools.
We have shown that, in a MIMO setting, the selection of
the best relay is suboptimal and large performance improvements
can be obtained by selecting the best antennas
distributed over multiple relays.
Numerical simulations have shown that
the proposed joint antenna-and-relay
selection approach significantly outperforms 
existing schemes, which exploit 
the same amount of P-CSI.


\appendices

\section{Proof of Lemma~1}
\label{app:prop2}

Preliminarily, we remember that $\C=\G  \,\F  \,\H  \,\F_0$
is full-column rank if {\bf a1)} and {\bf a2)} hold.
It can be shown (see, e.g., \cite{Li.2006}) that, conditioned on $\H$,  the $k$th diagonal entry
$\{ (\C^\herm  \C )^{-1}\}_{kk}$ of the matrix $(\C^\herm  \C )^{-1}$ follows an inverse-Gamma distribution, with shape parameter $\alpha \eqdef \Nd-\Nb+1$ and
scale parameter $\beta_k \eqdef {1}/{\{\R^{-1}\}_{kk}}$, where
$\R$
is defined in the lemma statement.
Thus, the probability density function of the random variable
$\{ (\C^\herm  \C )^{-1}\}_{kk}$, given $\H$, reads as
\be
p_k(x) = \frac{1}{\Gamma(\alpha)  \, \beta_k^{\alpha}}  \, x^{-\alpha-1}
\, e^{-\frac{1}{x  \beta_k}}
\label{eq:pdf}
\ee
where the gamma function $\Gamma(\alpha) = (\alpha-1)!$
since $\Nd-\Nb$ is a non-negative integer number \cite{Abramowitz.book.1965}.
Therefore, one has
\begin{multline}
\Es_{\G} \left [ \trace \left(\C^\herm  \, \C \right)^{-1}  \, \big | \, \H \right] =
\sum_{k=1}^{\Nb} \Es_{\G} \left[ \left\{ \left(\C^\herm  \, \C \right)^{-1}\right\}_{kk}
\, \big |\, \H \right]  \\
= \frac{1}{\Gamma(\alpha)} \sum_{k=1}^{\Nb}\frac{1}{\beta_k^{\alpha}}  \left(\lim_{\delta \rightarrow 0} \int_{\delta}^{+\infty} x^{-\alpha}   e^{-\frac{1}{x  \beta_k}}  \mathrm{d}x \right) \: .
\label{eq:prop2}
\end{multline}
After some calculations, eq.~\eqref{eq:prop2} can be rewritten as
\begin{multline}
\Es_{\G} \left [ \trace \left(\C^\herm  \, \C \right)^{-1}  \,\big | \, \H \right]  = \sum_{k=1}^{\Nb} \frac{\beta^{-1}_k}{\Gamma(\alpha)}
\lim_{\delta \rightarrow 0} \gamma(\alpha-1, (\delta  \, \beta_k)^{-1}) \\
= \frac{\Gamma(\alpha-1)}{\Gamma(\alpha)} \sum_{k=1}^{\Nb}{\beta^{-1}_k}   = \frac{1}{\alpha-1} \sum_{k=1}^{\Nb} \{\R^{-1}\}_{kk}
= \frac{\trace (\R^{-1} ) }{\Nd-\Nb}
\label{eq:media}
\end{multline}
where we have exploited the definition  of the incomplete gamma function
$\gamma(s, x) \eqdef \int_0^x t^{s-1}  e^{-t}  \mathrm{d}t$ \cite{Abramowitz.book.1965} and its asymptotic property
$\Gamma(s) = \lim_{x \rightarrow +\infty} \gamma(s, x)$.

\section{Hessian of the cost function~(15)}
\label{app:2}

Let us check convexity of a generic summand of the cost function
\eqref{eq:costmale}. To this end, it is sufficient to study the multivariate
function
\be
f(x, y_1, y_2, \ldots, y_n) \eqdef \frac{1}{A \, x \, (y_1 + y_2 + \cdots + y_n)}
\label{app:f}
\ee
with $A>0$, $x>0$, and $y_i >0$, for each $i \in \{1,2,\ldots, n\}$.
The domain of $f$ is therefore given by $\Rset_{+}^{n+1}$, which is a convex set.
The function $f$ is twice differentiable over its domain.
It is noteworthy that, when $n=1$, the function \eqref{app:f} ends up to
a generic summand of \eqref{eq:opt-1} or \eqref{eq:opt-2}.

Let us calculate the Hessian matrix $\triangledown^2 f \in \Rset^{(n+1) \times (n+1)}$,
whose entries are the second-order partial derivatives of $f$ at
$(x, y_1, y_2, \ldots, y_n) \in \Rset_{+}^{n+1}$, i.e.,
\be
\{\triangledown^2 f\}_{ij}  =
\begin{cases}
\frac{\partial^2}{\partial x^2} f, &
\text{for $i=j=1$} \: ; \\
\frac{\partial^2}{\partial x \, \partial y_j} f, &
\text{for $i=1$ and $j \in \{2,3, \ldots,n\}$ } \\\
& \text{for $j=1$ and $i \in \{2,3, \ldots,n\}$} \: ; \\
\frac{\partial^2}{\partial y_i \, \partial y_j} f, &
\text{for $i,j \in \{2,3, \ldots,n\}$} \: . \\
\end{cases}
\ee
We recall that the function $f$ is convex [concave] if and only if the Hessian matrix
$\triangledown^2 f$ is  positive [negative] semidefinite for all the points
belonging  to its domain.

Using standard calculus concepts, it can be verified that
\barr
\frac{\partial^2}{\partial x^2} f & = \frac{2}{A \, x^3 \, (y_1 + y_2 + \cdots + y_n)}
\\
\frac{\partial^2}{\partial x \, \partial y_j} f & =
\frac{1}{A \, x^2 \, (y_1 + y_2 + \cdots + y_n)^2}
\\
\frac{\partial^2}{\partial y_i \, \partial y_j} f & =
\frac{2}{A \, x \, (y_1 + y_2 + \cdots + y_n)^3} \: .
\earr
We note that all the entries of $\triangledown^2 f$ are nonnegative on $\Rset_{+}^{n+1}$.
In the particular case of $n=1$, it is readily seen that the determinant of
$\triangledown^2 f \in \Rset^{2 \times 2}$ is given by
\be
\det(\triangledown^2 f )= \frac{3}{A^2 \, x^4 \, y_1^4} >0
\ee
which shows that, when $n=1$, $f$ is a strictly convex function on $\Rset_{+}^{2}$. Therefore, since the sum of convex functions is convex \cite{Boyd.book.2004}, the cost functions
\eqref{eq:opt-1} or \eqref{eq:opt-2} are convex.

On the other hand, when $n >1$, by resorting to
the Laplacian determinant expansion by minors, it results that
\be
\det(\triangledown^2 f )= \sum_{j=1}^{n+1} (-1)^{j+1} \, \{\triangledown^2 f\}_{1j}
\, {\bm M}_{1 j}
\ee
where ${\bm M}_{1 j} \in \Rset^{n \times n}$ is a so-called minor of $\triangledown^2 f$, obtained by taking the determinant of $\triangledown^2 f$ with row $1$ and
column $j$ crossed out. It can be verified that ${\bm M}_{1 j}$ is zero,
for each $j \in \{1,2,\ldots, n+1\}$. Thus, the determinant of $\triangledown^2 f$
is zero at each point belonging to the domain of $f$ if $n >1$. This is sufficient
to infer that $\triangledown^2 f $ is neither positive nor negative
definitive, which implies in its turn that, when $n>1$, $f$ is neither strictly convex
nor strictly concave on $\Rset_{+}^{n+1}$.



\clearpage

\begin{IEEEbiography}
[{\includegraphics[width=1in,height=1.25in,clip,keepaspectratio]{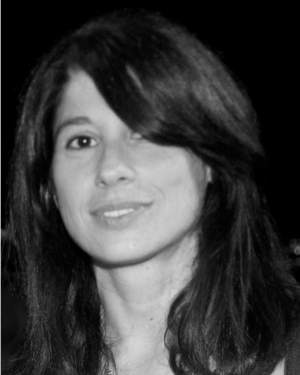}}]
{Donatella Darsena} (M'06-SM'16) received the Dr. Eng. degree summa cum laude in telecommunications engineering in 2001, and the Ph.D. degree in electronic and telecommunications engineering in 2005, both from the University of Napoli Federico II, Italy. From 2001 to 2002, she worked as embedded system designer in the Telecommunications, Peripherals and Automotive Group, STMicroelectronics, Milano, Italy. Since 2005, she has been an Assistant Professor with the Department of Engineering, University of Napoli Parthenope, Italy. Her research interests are in the broad area of signal processing for communications, with current emphasis on multicarrier modulation systems, space-time techniques for cooperative and cognitive communications, green communications for IoT. Dr. Darsena has served as an Associate Editor for the IEEE COMMUNICATIONS LETTERS from December 2016 to July 2019. 
Since August 2019 she has been a Senior Area Editor for IEEE COMMUNICATIONS LETTERS and Associate Editor for IEEE ACCESS since October 2018.
\end{IEEEbiography}

\vspace*{-2\baselineskip}

\begin{IEEEbiography}
[{\includegraphics[width=1in,height=1.25in,clip,keepaspectratio]{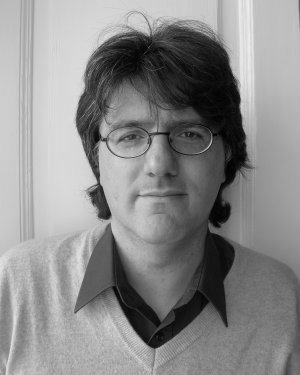}}]
{Giacinto Gelli}(M'18-SM'20)
was born in Napoli, Italy, on July 29, 1964.
He received the Dr. Eng. degree \textit{summa cum laude} in electronic
engineering in 1990, and the Ph.D. degree in computer science and
electronic engineering in 1994, both from the University of Napoli
Federico II.

From 1994 to 1998, he was an Assistant Professor with the
Department of Information Engineering, Second University of
Napoli.
Since 1998 he has been with the Department of Electrical Engineering and Information Technology, University of Napoli Federico II,
first as an Associate Professor,
and since November 2006 as a Full Professor of Telecommunications.
He also held teaching positions at the University Parthenope of
Napoli.
His research interests are in the broad area of
signal and array processing for communications,
with current emphasis on multicarrier modulation systems and
space-time techniques for cooperative and cognitive
communications systems.
 \end{IEEEbiography}

\vspace*{-2\baselineskip}

\begin{IEEEbiography}[
{\includegraphics[width=1in,height=1.25in,clip,keepaspectratio]{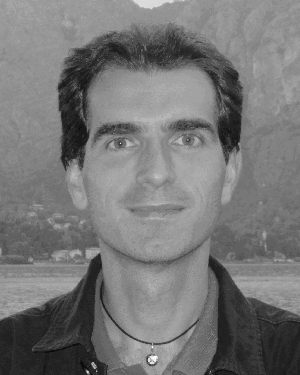}}]
{Francesco Verde}(M'10-SM'14) was born in Santa Maria Capua Vetere,
Italy, on June 12, 1974. He received the Dr. Eng. degree
\textit{summa cum laude} in electronic engineering
from the Second University of Napoli, Italy, in 1998, and the Ph.D.
degree in information engineering
from the University of Napoli Federico II, in 2002.
Since December 2002, he has been with the University of Napoli Federico II. He first served as an Assistant Professor of signal theory and mobile communications
and, since December 2011, he has served as an Associate Professor of telecommunications with the Department of Electrical Engineering and Information Technology.
His research activities include orthogonal/non-orthogonal multiple-access techniques, space-time processing for cooperative/cognitive communications, wireless systems optimization, and software-defined networks.

Prof. Verde has been involved in several technical program committees of major IEEE conferences in signal processing and wireless communications.
He has served as Associate Editor for IEEE TRANSACTIONS ON COMMUNICATIONS since 2017 and Senior Area Editor of the IEEE SIGNAL PROCESSING LETTERS since 2018. He was an Associate Editor of the IEEE TRANSACTIONS ON SIGNAL PROCESSING (from 2010 to 2014) and  IEEE SIGNAL PROCESSING LETTERS (from 2014 to 2018), as well as Guest Editor of the EURASIP Journal on Advances in Signal Processing in 2010 and SENSORS MDPI in 2018. 
\end{IEEEbiography}

\EOD
\end{document}